\documentstyle[12pt]{article}

\font\twelve=cmbx10 at 15pt
\font\ten=cmbx10 at 12pt

\def\build#1_#2^#3{\mathrel{\mathop{\kern 0pt#1}\limits_{#2}^{#3}}}

\def\vect#1{\overrightarrow{#1\kern 1pt}\kern-1pt}

\newcommand{\rf}[1]{(\ref{#1})}

\newcommand{\beq}[1]{\begin{equation}\label{#1}}
\newcommand{\eeq}{\end{equation}}
\newcommand{\beqa}[1]{\begin{eqnarray}\label{#1}}
\newcommand{\eeqa}{\end{eqnarray}}
\newcommand{\beqan}{\begin{eqnarray*}}
\newcommand{\eeqan}{\end{eqnarray*}}

\def\wt{\widetilde}

\def\ol{\overline}

\def\tg{\mathop{\rm tg}\nolimits}
\def\cotg{\mathop{\rm cotg}\nolimits}

\begin{document}

\begin{titlepage}

\begin{center}

\renewcommand{\thefootnote}{\fnsymbol{footnote}}

{\ten Centre de Physique Th\'eorique\footnote{
Unit\'e Propre de Recherche 7061} - CNRS - Luminy, Case 907}
{\ten F-13288 Marseille Cedex 9 - France }

\vspace{2 cm}

{\twelve LATEST DEVELOPMENTS ON THE BESS MODEL}

\vspace{0.3 cm}

{\bf Pierre CHIAPPETTA}

\vspace{4,2 cm}

{\bf Abstract}

\end{center}

The idea of a strongly interacting sector as responsible for
the electroweak symmetry breaking is tested through an effective
lagrangian description, called the BESS model, constructed on the
standing point of custodial symmetry and gauge invariance, without
specifing any dynamical scheme.

\vspace{4,2 cm}


\bigskip

\renewcommand{\thefootnote}{\fnsymbol{footnote}}

\noindent Number of figures : 4\footnote[7]{Figures can be obtained
via ordinary mail. Please ask them to : Secr\'etariat du CPT}

\bigskip

\noindent October 1994

\noindent CPT-94/P.3075

\bigskip

\noindent anonymous ftp or gopher: cpt.univ-mrs.fr

\end{titlepage}

\setcounter{footnote}{0}
\renewcommand{\thefootnote}{\arabic{footnote}}

\section{Introduction}

Although the standard model of electroweak interactions (hereafter
denoted as SM) is in perfect agreement with experimental data ---
at a precision level of below the 0.5\% --- the mechanism of
electroweak symmetry breaking (hereafter denoted as EWSB) is
still an open question and certainly the central issue and the main
goal for future accelerators.

In our opinion one possibility is a perturbative scheme with a light
Higgs which has still to be discovered. It needs a new symmetry
like supersymmetry to be viable. Otherwise, as shown by M.~Consoli
in this workshop~\cite{1}, the strong evidence for triviality of
$\lambda \phi^4$ theories leads to a non perturbative scheme without
self interacting scalar sector and with a heavy Higgs of mass $M_H^2
= 8 \pi^2 v^2$, $v$ being the vacuum expectation value of Higgs
field.

In what follows we will assume that the fundamental theory of
electroweak interactions is not precisely known, but that the
possible symmetries are, i.e. gauge invariance and custodial symmetry.

We are interested in performing a spontaneous symmetry breaking
avoiding physical scalar particles, i.e. by a non linear
realisation. We will first explicitely show that the SM appears
as a gauged $S U (2)_L \otimes S U (2)_R$ non linear
$\sigma$-model~\cite{2}.
Since this theory is non renormalisable it corresponds to an
effective one. The use of effective Lagrangians dates from beggining
of $60$'s with the introduction of the non linear $\sigma$-model as
an effective theory for low energy strong interactions, exhibiting a
chiral symmetry breakdown~\cite{3}.

In what follows we will extensively use the fact that any non linear
$\sigma$-model is gauge equivalent to theories with additional
hidden local symmetry~\cite{4}. This has been shown in the context
of pion interactions successfull to describe the $\rho$ vector
resonances, which correspond to the gauge bosons associated to this
additional hidden symmetry. Applying this mechanism to weak
interactions we will build vector and axial vector resonances as
gauge bosons associated to the hidden symmetry group of $S U (2)$
type. Under the assumption they are dynamical, we will get the $SU
(2)$ BESS Lagrangian~\cite{5} (BESS standing for Breaking
Electroweak Symmetry Strongly).

The existence of these new bosons called $\smash{\vect V}$ will
indirectly manifest at LEP through deviations from SM
expectations~\cite{6}. For this purpose a low energy effective theory
valid for heavy resonances will be derived.

The new particles are naturally coupled to fermions through mixing
between $\smash{\vect W}$ and $\smash{\vect V}$, although a direct
coupling is possible. Deviations from SM trilinear and quadrilinear
gauge bosons couplings are expected.

In the vector dominance approximation the BESS model corresponds to
technicolor model~\cite{7} with a single technidoublet.
In order to reproduce the one family technicolor model~\cite{8} one
has to extend the original $S U (2)_L \otimes S U (2)_R$ symmetry
group to $S U (8)_L \otimes S U (8)_R$. This leads to the extended
BESS model and to the existence of pseudogoldstone bosons~\cite{9}.

These models are already constrained from the precision electroweak
data as we will see. Exploration of the usefulness of very
energetic linear $e^+ e^-$ colliders and LHC to detect the new
particle spectrum will also be reviewed.

\section{Standard model revisited as a gauged non linear
$\sigma$-model}

The non fermionic part of the SM is described by the Lagrangian:
\beqa{1}
& {\cal L}_{WS} = \frac{1}{2 g'^2} Tr
\left( B_{\mu \nu} B^{\mu \nu} \right)
& + \frac{1}{2 g^2} Tr
\left( F_{\mu \nu} (W) F^{\mu \nu} (W) \right)
+ D_{\mu} \phi^+ D^{\mu} \phi \nonumber \\[1mm]
& & - \lambda
\left( \phi^+ \phi + \frac{\mu^2}{2 \lambda} \right)^2
\eeqa

\noindent where $B_{\mu \nu}$ (resp $F_{\mu \nu}$) is the field
strength of the $U (1)$ (resp $S U (2)$) field and $\phi =
\left( {\phi^+ \atop \phi^0} \right)$ the Higgs doublet.
\beqa{2}
B_{\mu \nu}
& = & \partial_{\mu} B_{\nu} - \partial_{\nu} B_{\mu}
\quad \mbox{where} \quad B_{\mu} = i g'\tau_3
\frac{B'_{\mu}}{2} \nonumber \\[1mm]
W_{\mu \nu}
& = & \partial_{\mu} W_{\nu} - \partial_{\nu} W_{\mu} +
[W_{\mu}, W_{\nu}]
\quad \mbox{where} \quad W_{\mu} = i g
\frac{\vec{\tau}}{2} \cdotp \vect W'_{\mu}
\eeqa

\noindent $D_{\mu} = \partial_{\mu} + W_{\mu} + B_{\mu}$ is the
covariant derivative.

Let us assume $g = g' = 0$. The scalar part of~\rf{1} reads:
\beq{3}
{\cal L}_S = \partial_{\mu} \phi^+ \partial^{\mu} \phi - \lambda
\left( \phi^+ \phi + \frac{\mu^2}{\lambda} \right)^2
\eeq
exhibiting a global $S U (2)_L \otimes S U (2)_R$ global symmetry.
Indeed introducing a matricial notation:
\beq{4}
M = \sqrt{2} \left( i \tau_2 \phi^{\star},\ \phi \right)
\eeq
the expression~\rf{3} becomes:
\beq{5}
{\cal L}_S = \frac{1}{4} Tr
\left( \partial_{\mu} M^+ \partial^{\mu} M \right)
- \frac{\lambda}{4}
\left( \frac{1}{2} Tr (M^+ M) + \frac{\mu^2}{2 \lambda} \right)^2
\eeq
invariant under
\beq{6}
M \to M' = L M R^+ = e^{
i \vect{\varepsilon}\!\!\!_L \cdot \vect \tau \! / 2
}
M e^{
- i \vect{\varepsilon}\!\!_R \cdot \vect \tau \! / 2
}
\eeq

\noindent If one assume $\mu^2 < 0$ the potential has a minimum for
\beq{7}
M^+ M = - \frac{\mu^2}{\lambda} \equiv v^2
\eeq
leading to a spontaneous breaking of $S U (2)_L \otimes S U (2)_R$
into $S U (2)_D$.

The scalar part of the Lagrangian becomes:
\beq{8}
{\cal L}_S = \frac{1}{4} Tr \partial_{\mu} M^+
\partial^{\mu} M - \frac{\lambda}{4}
\left( \frac{1}{2} Tr M^+ M - v^2 \right)^2
\eeq
Using
\beq{9}
M = \sigma + i \vect{\pi} \cdot \vect \tau
\eeq
the standard model is a linear $\sigma$ model
\beq{10}
{\cal L}_S
= \frac{1}{2} \partial_{\mu} \sigma \partial^{\mu} \sigma
+ \frac{1}{2} \partial_{\mu} \vect{\pi} \partial^{\mu} \vect{\pi}
- \frac{\lambda}{4} \left( \sigma^2 + \vect{\pi}^2 - v^2 \right)^2
\eeq
where $\sigma$ is the Higgs field.

Let us take now the limit $\lambda \to \infty$. The generating
functionnal~\cite{10}~:
\beqa{11}
& G [\lambda] = \int [d \vect{\pi}] [d \sigma] & e^{
i \int d^4 x
\left(
\frac{1}{2} \partial_{\mu} \sigma \partial^{\mu} \sigma +
\frac{1}{2} \partial_{\mu} \vect{\pi}
\partial^{\mu} \vect{\pi}
\right)
} \nonumber \\[1mm]
& & e^{
- i \int d^4 x \frac{\lambda}{4}
\left( \sigma^2 + \vect{\pi}^2 - v^2 \right)^2
}
\eeqa
becomes
\beq{12}
G [\lambda] \build{=}_{\lambda \to \infty}^{}
\int [d \vect{\pi}] [d \sigma] \ \delta
\left( \sigma^2 + \vect{\pi}^2 - v^2 \right) e^{
i \int d^4 x
\left(
 \frac{1}{2} \partial_{\mu} \sigma \partial^{\mu} \sigma
+ \frac{1}{2} \partial_{\mu} \vect{\pi} \partial^{\mu} \vect{\pi}
\right)
}
\eeq
leading to the non linear $\sigma$ model Lagrangian:
\beq{13}
{\cal L}_S = \frac{1}{2} \partial_{\mu} \vect{\pi} \partial^{\mu}
\vect{\pi} + \frac{1}{2} \frac{
\left( \vect{\pi} \partial_{\mu} \vect{\pi} \right)
\left( \vect{\pi} \partial^{\mu} \vect{\pi} \right)
}{v^2 - \vect{\pi}^2}
\eeq
which can be rewritten as:
\beq{14}
{\cal L}_S = \frac{v^2}{4} Tr
\left( \partial_{\mu} U \partial^{\mu} U^+ \right)
\eeq
where $U$ is an unitary matrix.

We introduce the Goldstone bosons $\varphi$ by parametrizing $U$ as:
\beq{15}
U = e^{
i \vect{\varphi} (x) \cdot \vect \tau \! / v
}
\eeq
Adding now the bosonic fields, the SM is a gauged non linear $\sigma$
model:
\beq{16}
{\cal L}_B = \frac{v^2}{4} Tr
\left( D_{\mu} U D^{\mu} U^+ \right) + \frac{1}{2 g'^2} Tr
\left( B_{\mu \nu} B^{\mu \nu} \right) + \frac{1}{2 g^2} Tr
\left( F_{\mu \nu} (W) F^{\mu \nu} (W) \right)
\eeq
Going into unitary gauge $U = 1$ one gets the mass terms for gauge
bosons.

\section{Hidden gauge symmetry}

The basic idea of hidden gauge symmetry is that any non linear
$\sigma$ model defined on coset space $G / H$ is gauge equivalent to
$G \otimes H$ local where $H$ local is the hidden gauge symmetry
group~\cite{11}.

Let us write
\beq{17}
U = L R^+
\eeq

\noindent $L (x) \in S U (2)_L$ and $R (x) \in S U (2)_R$ are
the Goldstone bosons. We define:
\beq{18}
\rho (x) =
\normalbaselineskip=18pt
\pmatrix{
L (x) & 0     \cr
0     & R (x) \cr
}
\eeq
and introduce the Maurer Cartan form:
\beq{19}
\omega_{\mu} (x) = \rho^+ (x) \partial_{\mu} \rho (x) =
\normalbaselineskip=18pt
\pmatrix{
L^+ \partial_{\mu} L & 0                    \cr
0                    & R^+ \partial_{\mu} R \cr
}
\eeq
We split into:
\beq{20}
\omega''_{\mu} = \frac{1}{2}
\normalbaselineskip=18pt
\pmatrix{
L^+ \partial_{\mu} L + R^+ \partial_{\mu} R & 0 \cr
0 & L^+ \partial_{\mu} L + R^+ \partial_{\mu} R \cr
}
\eeq
and
\beq{21}
\omega_{\mu}^{\bot} = \frac{1}{2}
\normalbaselineskip=18pt
\pmatrix{
L^+ \partial_{\mu} L - R^+ \partial_{\mu} R & 0 \cr
0 & - \left( L^+ \partial_{\mu} L - R^+ \partial_{\mu} R \right) \cr
}
\eeq
Under local $S U (2)_V$ transformations $h (x)$ they behave as:
\beq{22}
L^+ \partial_{\mu} L \pm R^+ \partial_{\mu} R \to
h^+
\left( L^+ \partial_{\mu} L \pm R^+ \partial_{\mu} R \right)
h +
\left( h^+ \partial_{\mu} h \pm h^+ \partial_{\mu} h \right)
\eeq
i.e.,
\beq{23}
\omega_{\mu}^{\bot} \to \wt h^+ \omega_{\mu}^{\bot} \wt h
\quad \mbox{and} \quad
\omega''_{\mu} \to \wt h^+ \omega''_{\mu} \wt h +
\wt h^+ \partial_{\mu} \wt h
\eeq
with $\wt h =
\normalbaselineskip=18pt
\pmatrix{
h & 0 \cr
0 & h \cr
}
$

$\omega''_{\mu}$ transforms under $S U (2)_V$ as a gauge bosons
${\cal V}_{\mu}$:
\beq{24}
{\cal V}_{\mu} \to h^+ {\cal V}_{\mu} h + h^+ \partial_{\mu} h
\eeq
where ${\cal V}_{\mu} = i g_V \frac{\smash{\vect {\tau}}}{2} \cdot
V'_{\mu}$.

The most general Lagrangian one gets from~\rf{19} is:
\beq{25}
{\cal L} = - \frac{v^2}{2}
\left[
Tr \left( \omega_{\mu}^{\bot} \omega^{\bot \mu} \right) +
\alpha \ Tr
\left(
\left( \omega''_{\mu} - \wt{\cal V}_{\mu} \right)
\left( \omega''^{\mu} - \wt{\cal V}^{\mu} \right)
\right)
\right]
\eeq
where $\wt{\cal V}_{\mu} =
\normalbaselineskip=18pt
\pmatrix{
{\cal V}_{\mu} & 0              \cr
             0 & {\cal V}_{\mu} \cr
}
$ and $\alpha$ is an arbitrary parameter.

The SM corresponds to the first term of~\rf{25} and $\wt{\cal
V}_{\mu} = \omega''_{\mu}$ is an auxiliary field.

In order to enable these gauge bosons to show up as physical
particles we have to add to the previous Lagrangian a kinetic term:
\beq{26}
{\cal L}_{KIN} = \frac{1}{2 g_V^2} Tr
\left( F_{\mu \nu} ({\cal V}) F^{\mu \nu} ({\cal V}) \right)
\eeq

\noindent This term can be generated for $2 D$ and $3 D$
theories~\cite{12} and from quantum corrections for $4 D$
theories~\cite{13}. To get the SM we have to perform a gauging of $S
U (2)_L \otimes S U (2)_R$ into $S U (2)_L \otimes U (1)_Y$ and add
kinetic terms for $W$ and $B$ fields (see eq.~\rf{1}). One
gets~\cite{14} a Yang Mills Lagrangian whose gauge group is $S U (2)_L
\otimes U (1)_Y \otimes S U (2)_V$ after replacement of the
derivatives by the covariant ones:
$$
D_{\mu} L =
\left( \partial_{\mu} + W_{\mu} - {\cal V}_{\mu} \right) L
$$
and
$$
D_{\mu} R =
\left( \partial_{\mu} + B_{\mu} - {\cal V}_{\mu} \right) R .
$$

\noindent In the unitary gauge $(L = 1,\ R = 1)$ one gets:
\beqa{27}
& {\cal L}_{BESS} = & - \frac{v^2}{4}
\left[ Tr (W - B)^2 + \alpha \ Tr (W + B - V)^2 \right]
\nonumber \\[1mm]
& &
+ \frac{1}{2 g^2} Tr
\left( F_{\mu \nu} (W) F^{\mu \nu} (W) \right)
+ \frac{1}{2 g'^2} Tr
\left( B_{\mu \nu} B^{\mu \nu} \right)
\nonumber \\[1mm]
& &
+ \frac{1}{2 g''^2} Tr
\left( F_{\mu \nu} (V) F^{\mu \nu} (V) \right)
\eeqa
with $F_{\mu \nu} (V) = \partial_{\mu} V_{\nu} -
\partial_{\nu} V_{\mu} +
\frac{1}{2} [V_{\mu},\ V_{\nu}]$ where $V_{\mu} = i g''
\frac{\smash{\vect {\tau}}}{2} \cdot \smash{\vect V'_{\mu}}$
after the rescaling $g_V \equiv \frac{g''}{2} \cdotp$ The first term
within brackets is the usual mass term appearing the SM.

The fermionic part of the Lagrangian reads:
\beqa{28}
& {\cal L}_F = & i \ol{\psi}_L \gamma^{\mu}
\left(
\partial_{\mu} + W_{\mu} + i
\left( Q - \frac{\tau_3}{2} \right)
g' B'_{\mu}
\right)
\psi_L \nonumber \\[1mm]
& &
+ i \ol{\psi}_R \gamma^{\mu}
\left(
\partial_{\mu} + i Q g' B'_{\mu}
\right)
\psi_R \nonumber \\[1mm]
& &
+ i b \ol{\psi}_L \gamma^{\mu}
\left(
\partial_{\mu} + \frac{V_{\mu}}{2} + i
\left( Q - \frac{\tau_3}{2} \right)
g' B'_{\mu}
\right)
\psi_L
\eeqa
where the last term corresponds to a direct coupling of the fermions
to the fields $V$ and $Q$ is the electric charge.

The physical
vector bosons $W^{\pm}, V^{\pm}, A, Z^0$ and $V^0$ are
obtained after diagonalization of the charged and neutral
sectors~\cite{14}.
The mixing angles are of the order of $0
\left( \frac{g}{g''} \right)$.

We will now study the low energy effects of the model.

\section{Low energy effects of vector resonances}

We will evaluate~\cite{15} the solution of the classical equations of
motion in the limit $M_V \to \infty$.

{}From
\beq{29}
\frac{\partial}{\partial V_{\mu}^a}
\left( {\cal L}_{BESS} + {\cal L}_F \right) = 0
\eeq
we get:
\beq{30}
g'' V_{\mu}^a
= g W_{\mu}^a
+ g' B'_{\mu} \delta_{a_3}
+ \frac{b}{v^2 \alpha}
\ol{\psi}_L \gamma^{\mu} \psi_L \tau_a
\eeq

\noindent The last term in eq.~\rf{30} will be neglected since $b$
is small.

The physical effect of $V$ to low energy is present in the $V$
kinetic term and the interacting fermionic Lagrangian where one has
to replace $V_{\mu}$ by expression~\rf{30}.

Neglecting for the moment the trilinear and quadrilinear couplings
one gets:
\beqa{31}
& {\cal L}_{eff} =
& - \frac{1}{4} (1 + Z_{\gamma}) A_{\mu \nu} A^{\mu \nu}
- \frac{1}{2} (1 + Z_W) W_{\mu \nu}^+ W^{\mu \nu -}
\nonumber \\[1mm]
& & - \frac{1}{4} (1 + Z_Z) Z_{\mu \nu} Z^{\mu \nu}
+ \frac{1}{2} Z_{Z_{\gamma}} A_{\mu \nu} Z^{\mu \nu}
\nonumber \\[1mm]
& & - M_W^2 W_{\mu}^+ W^{\mu -}
- \frac{1}{2} M_Z^2 Z_{\mu} Z^{\mu}
\eeqa
with:
\beqa{32}
Z_{\gamma} & = & \left( \frac{g}{g''} \right)^2
4 \sin^2 \theta_W \nonumber \\[1mm]
Z_W & = & \left( \frac{g}{g''} \right)^2
\nonumber \\[1mm]
Z_Z & = & \left( \frac{g}{g''} \right)^2
\frac{\cos^2 2 \theta_W}{\cos^2 \theta_W} \nonumber \\[1mm]
Z_{Z_{\gamma}} & = & - \left( \frac{g}{g''} \right)^2
2 \tg \theta_W \cos 2 \theta_W
\eeqa

\noindent The corrections to the SM produce a wave function
renormalization of the fields $A_{\mu}, Z_{\mu}$ and $W^{\pm}_{\mu}$
and a mixing term $Z_{Z \gamma}$ to be cancelled.

Defining new fields indiced by the superscript $R$:
\beqa{33}
A_{\mu} & \to & \left( 1 - \frac{Z_{\gamma}}{2} \right)
A_{\mu}^R + Z_{Z_{\gamma}} Z_{\mu}^R \nonumber \\[1mm]
W_{\mu} & \to & \left( 1 - \frac{Z_W}{2} \right)
W_{\mu}^R \nonumber \\[1mm]
Z_{\mu} & \to & \left( 1 - \frac{Z_Z}{2} \right)
Z_{\mu}^R
\eeqa
the only deviations are present in the mass terms of eq.~\rf{31} i.e.
\beq{34}
M_W^2 \to M_W^2 (1 - Z_W) \quad \mbox{and} \quad
M_Z^2 \to M_Z^2 (1 - Z_Z).
\eeq

The couplings of the gauge bosons to fermions will also be affected:
\beq{35}
{\cal L}_{em} = - e \left( 1 - \frac{Z_{\gamma}}{2} \right)
\ol{\psi} \gamma^{\mu} Q \psi A_{\mu}^R
\eeq
\beq{36}
{\cal L}_{charged} = - \frac{e}{\sqrt{2} \sin \theta_W}
W_{\mu}^R \ol{\psi}_u \gamma^{\mu}
\frac{1 - \gamma_5}{2} \psi_d
\left( 1 - \frac{b}{2} - \frac{Z_W}{2} \right) + h.c.
\eeq
\beqa{37}
{\cal L}_{neutral} & = & - \frac{e}{\sin \theta_W \cos \theta_W}
\left( 1 - \frac{b}{2} - \frac{Z_Z}{2} \right)
\ol{\psi} \gamma^{\mu}
\left[
\vphantom{\frac{b}{2}} T_{3 L} \frac{1 - \gamma_5}{2}
\right.
\nonumber \\[1mm]
& &
\left.
- Q \sin^2 \theta_W
\left( 1 + \frac{b}{2} - \cotg \theta_W Z_{Z_{\gamma}} \right)
\right]
\psi Z_{\mu}^R
\eeqa

\noindent The input parameters that are used for LEP Physics are:

\begin{itemize}

\item[---] the electric charge $e^{ph}$

\item[---] the Fermi constant $G_F^{ph}$

\item[---] the mass of the $Z$ boson $M_Z^{ph}$

\end{itemize}

They are identified as follows:
\beq{38}
e^{ph} = e \left( 1 - \frac{Z_{\gamma}}{2} \right)
\eeq
\beq{39}
M_Z^{2 ph} = M_Z^2 \left( 1 - Z_Z \right)
\eeq
\beq{40}
\frac{G_F}{\sqrt{2}} = \frac{
e^{2 ph} \left( 1 - b - Z_Z + Z_{\gamma} \right)
}{
8 \sin^2 \theta_W \cos^2 \theta_W M_Z^{2 ph}
}
\eeq
Since
\beq{41}
\frac{G_F}{\sqrt{2}} = \frac{
4 \pi \alpha (M_Z)
}{
8 \sin^2 \theta_W^{SM} \cos^2 \theta_W^{SM} M_Z^{2 ph}
}
\eeq

\noindent we can connect $\sin \theta_W$ to the tree level SM
value $\sin \theta_W^{SM}$.

The strength of corrections due to $V$ to the ratio:
\beq{42}
\frac{M_W^2}{M_Z^2} = \frac{1}{2} + \sqrt{
\frac{1}{4} -
\frac{\pi \alpha (M_Z)}{\sqrt{2} G_F M_Z^2 (1 - \Delta r_W)}
}
\eeq
is given by
\beq{43}
\Delta r_W = 2 \left( \frac{g}{g''} \right)^2 - b
\eeq

The neutral Lagrangian~\rf{37} can be rewritten as:
\beq{44}
{\cal L}_{neutral} = -
\frac{e^{ph} Z_{\mu}}{\sin \theta_W^{SM} \cos \theta_W^{SM}}
\ol{\psi} \gamma^{\mu}
\left( g_V + \gamma^5 g_A \right) \psi
\eeq
with
\beqa{45}
g_V
& = & \frac{T_{3_L}}{2} - Q \sin^2 \ol{\theta}_W \nonumber \\[1mm]
g_A
& = & - \frac{T_{3_L}}{2} \nonumber \\[1mm]
\sin^2 \ol{\theta}_W
& = & (1 + \Delta K) \sin^2 \theta_W^{SM}
\eeqa
where
\beq{46}
\Delta K = \frac{1}{\cos 2 \theta_W^{SM}}
\left(
\left( \frac{g}{g''} \right)^2 - \frac{b}{2}
\right)
\eeq
We get also
\beq{47}
\Delta \rho = 0
\eeq

LEP 200 will directly test the non abelian gauge structure through
the trilinear and quadrilinear vertices among gauge bosons.

The anomalous vertices read:
\beqa{48}
{\cal L}_{trilinear}
& = & \frac{i e^{ph}}{\sin^2 \theta_W^{SM} \cos^2 \theta_W^{SM}}
\left(
b \cos^2 \theta_W^{SM} - \left( \frac{g}{g''} \right)^2
\right) \nonumber \\[1mm]
& &
\left(
Z_{\mu \nu} W^{- \mu} W^{+ \nu} +
Z^{\mu} W^{- \nu} W_{\mu \nu}^+ +
Z^{\nu} W^{+ \mu} W_{\mu \nu}^-
\right)
\eeqa
and
\beqa{49}
{\cal L}_{quadri}
& = &
\left(
2 g_{\mu \rho} g_{\nu \sigma} -
g_{\mu \nu} g_{\rho \sigma} -
g_{\mu \sigma} g_{\rho \nu}
\right)
W^{+ \mu} W^{- \rho} \nonumber \\[1mm]
& & \left[
- e^{2 ph} \cotg \theta_W^{SM} \gamma_1 A^{\nu} Z^{\sigma} +
\frac{1}{2} \frac{e^{2 ph}}{\sin^2 \theta_W^{SM}}
\gamma_2 W^{+ \nu} W^{- \sigma}
\right. \nonumber \\[1mm]
& & \left.
- \frac{1}{2}
e^{2 ph} \cotg^2 \theta_W^{SM} \gamma_3 Z^{\nu} Z^{\sigma}
\vphantom{\frac{e^{2 ph}}{\sin^2 \theta_W^{SM}}}
\right]
\eeqa
with
\beqan
\gamma_1
& = & \frac{1}{2 \cos^2 \theta_W^{SM} \cos 2 \theta_W^{SM}}
\left(
b \cos^2 \theta_W^{SM} - \left( \frac{g}{g''} \right)^2
\right) \\[1mm]
\gamma_2
& = & \frac{\cos^2 \theta_W^{SM}}{\cos 2 \theta_W^{SM}}
b + \left( \frac{g}{g''} \right)^2
\left(
\frac{1}{4} - \frac{1}{\cos 2 \theta_W^{SM}}
\right) \\[1mm]
\gamma_3
& = & \frac{b}{\cos 2 \theta_W^{SM}}
- \left( \frac{g}{g''} \right)^2
\frac{1 + 2 \cos^2 \theta_W^{SM}}{
4 \cos^4 \theta_W^{SM} \cos 2 \theta_W^{SM}
}
\eeqan
We are now ready to derive the bounds on the BESS parameter space
coming from precise LEP measurements and study the potential
discovery at future colliders.

\section{LEP constraints}

The analysis of LEP data, concerning the total width, the hadronic
width, the leptonic width, the leptonic and $b$ forward-backward
asymmetries, the $\tau$-polarization, the cesium atomic parity
violation and the ratio $M_W / M_Z$, uses available full one loop
radiative correction programs~\cite{16}. This brings in a dependance
on $\alpha_S,\ m_{top}$ and a cut off $\Lambda$ which corresponds to
the Higgs mass in the Standard Model. The quantities $\Delta K,\
\Delta
\rho,\ \Delta r_W$ are directly connected to observable quantities.
We will reexpress them in terms of the parameters
$\varepsilon_i$~\cite{17}~:
\beqa{50}
\varepsilon_1
& = & \Delta \rho \nonumber \\[1mm]
\varepsilon_2
& = & \cos^2 \theta_W \Delta \rho +
\frac{\sin^2 \theta_W}{\cos 2 \theta_W}
\Delta r_W - 2 \sin^2 \theta_W \Delta K \nonumber \\[1mm]
\varepsilon_3
& = & \cos^2 \theta_W \Delta \rho +
\cos 2 \theta_W \Delta K
\eeqa

The BESS contribution reads:
\beqa{51}
\varepsilon_1
& = & \varepsilon_2 = 0 \nonumber \\[1mm]
\varepsilon_3
& = & \left( \frac{g}{g''} \right)^2 - \frac{b}{2}
\eeqa

This shows explicitely that through LEP data we are only sensitive
to one combination of BESS parameters i.e. $\varepsilon_3$. The
allowed region at $90\%$ CL in the $\left(
\left( \frac{g}{g''} \right)^2,\ \frac{b}{2}
\right)$ plane is shown in fig.~1 for three top mass values.

The chosen experimental value~\cite{18}
\beq{52}
\varepsilon_3^{\exp} = (3.4 \pm 1.8) 10^{-3}
\eeq
corresponds to La Thuile et Moriond data.

The two standard deviation from Standard Model expectation for $b$
partial width can be expressed in terms of $\varepsilon_b$
parameter~\cite{19}.

Assuming a non zero direct coupling only for the heaviest generation
(as expected from one loop BESS radiative corrections proportional
to $m_f$) we get:
\beq{53}
\varepsilon_b = - \frac{b}{2}
\eeq

After adding the SM expectation for $m_t = 170$ GeV we get at 90\%
CL
\beq{54}
- 3.0\ 10^{- 2} \le b \le - 3.4\ 10^{- 3}.
\eeq

\section{Discovery potential at future colliders}

Provided the center of mass energy is higher than $2 M_W$ we are
directly sensitive to trilinear and quadrilinear gauge bosons
couplings. LEP 200 energy is too small to be really sensitive to the
expected BESS model deviations. Fortunately the planned linear
$e^+ e^-$ colliders will put severe constraints especially
through the reaction $e^+ e^- \to W^+ W^-$ which deviates from SM
values due to $V^0$ exchange~\cite{20}. The best constraints are
obtained from longitudinally polarized $W_L^+ W_L^-$ final state as
shown in fig.~2. Needless to remind that if a $V^0$ resonance
exists below the center of mass energy it will show up in $e^+ e^-
\to f \ol f,\ W^+ W^-$.

 Hadronic colliders as especially the LHC are
well suited to discover the charged resonances $V^{\pm}$~\cite{21}.

Two subprocesses contribute: the quark antiquark annihilation and
the $\gamma W$ or $Z W$ fusion. The appropriate final state is $W
Z$ followed by leptonic decays which is not affected by the top
background since it can be reduced from $Z$ mass reconstruction of
lepton pair and use of isolation criteria for leptons~\cite{22}.

The charged resonances show up as broad resonances around the
$V^{\pm}$ mass in $W Z$ invariant mass spectrum or a broad Jacobian
peak around
$\frac{M_V}{2}$ in the $Z$ transverse momentum spectrum.

At LHC, assuming an integrated luminosity of $10^5 pb^{-1} /$year,
one can reach discovery limits up to 2 TeV. Nevertheless this limit
cannot be reached for the full left over $\left(
\frac{g}{g''},\ b
\right)$ parameter space. Fig.~3 shows the favorable choice $(g''
= 20,\ b = 0.0\ 16)$ for $M_V = 1.5$ TeV whereas as can be inferred
{}from fig.~4 for $g'' = 20$ and $b = 2.0\ 10^{-3}$ the signal
exhibits no singular behaviour from background.

\section{Extended BESS model}

One important specialization of BESS model is to technicolor
theories since for particular values of the parameters it would
correspond to a technicolor model invol\-ving a single technidoublet
$(N_d = 1)$. If a non zero direct coupling of $V$ to fermions exists
it corresponds to an extended technicolor~\cite{23} with:
\beq{55}
b = - 2 \left( \frac{v}{\Lambda_{ETC}} \right)^2
\eeq
where $\Lambda_{ETC}$ is the scale for extended technicolor.

Previous analyses have shown that a conventional QCD scaled
technicolor is excluded for $N_{TC} N_d \le 12$ at 90\% CL. Therefore
we will extend the original non linear $S U (2)_L \otimes S U (2)_R$
$\sigma$-model to a $S U (8)_L \otimes S U (8)_R$ broken to $S U
(8)_V$ in order to incorporate the one family technicolor model $(N_d
= 4)$. The model~\cite{9} will contain not only vector resonances but
also axial vector resonances and pseudogoldstone bosons.

The construction starts from a gauged non linear $\sigma$-model of
gauge group\linebreak $S U (3)_C \otimes S U (2)_L \otimes U (1)_Y$:
\beq{56}
{\cal L} = \frac{v^2}{16} Tr
\left[ (D_{\mu} U) D^{\mu} U^+ \right]
\eeq
with
$$
U = \exp \left( \frac{2 i \pi^A \cdot T^A}{v} \right)
$$

The $\pi^A$ are the Goldstone bosons whereas the $T^A$ are the $S U
(8)$ generators:
$$
T^A = \left(
T^a,\ \wt T^a,\ T^D,\ T_8^{\alpha},\
T_8^{a \alpha},\ T_3^{\mu i},\ \wt T_3^{\mu i}
\right)
$$
where $A = 1, \cdots 63,\ a = 1,2,3$ is a $S U (2)$ triplet index,
$\alpha = 1, \cdots 8$ corresponds to $S U (3)_C$ octet indices, $i =
1,2,3$ to $S U (3)_C$ triplet indices and $\mu = (0, a)$. $T^D$ is a
singlet under $S U (3)_C \otimes S U (2)_L \otimes U (1)_Y$.

The SM gauge fields can be written as:
\beqa{57}
{\cal A}_{\mu}
& = & 2 i g W_{\mu}^a T^a + i \sqrt{2} g_S G_{\mu}^{\alpha}
T_8^{\alpha} + 2 i \frac{1}{\sqrt{3}} g' B_{\mu} T^D
\nonumber \\[1mm]
{\cal B}_{\mu}
& = & 2 i g' B_{\mu} \left( T^3 + \frac{1}{\sqrt{3}} T^D \right)
+ i \sqrt{2} g_S G_{\mu}^{\alpha} T_8^{\alpha}
\eeqa
and the new resonances:
\beqan
V_{\mu} & = & i g'' V_{\mu}^A T^A \\[1mm]
A_{\mu} & = & i g'' A_{\mu}^A T^A
\eeqan

The most general Lagrangian for gauge bosons reads:
\beqa{58}
{\cal L}
& = & - \frac{v^2}{16} \left(
a\ Tr ({\cal A} - {\cal B})^2 + b\ Tr ({\cal A} + {\cal B} - 2 V)^2
\right.
\nonumber \\[1mm]
& & + \left.c
Tr ({\cal A} - {\cal B} + 2 A)^2 + d\ Tr (2 A)^2
\right)
\nonumber \\[1mm]
& & + \ \mbox{kinetic terms}
\eeqa

The $A$ mass is a new parameter and we will assume no
direct coupling to fermions.

The pseudo-Goldstone mass spectrum has been derived~\cite{24} from
the one loop effective potential, which includes, besides the
ordinary gauge interactions, the Yukawa couplings. The masses are
proportional to the ultraviolet cut-off $\Lambda$ and depending on
the heaviest fermions masses.

The restrictions from LEP precision measurements are obtained from
the $\varepsilon_i$ parameters. The parameters $\varepsilon_1$ and
$\varepsilon_2$, which are isospin violating, are sensitive to the
pseudogoldstone mass spectrum. The effect on the parameter
$\varepsilon_1$ is to weaken the upper bound on the top mass. The
contribution to $\varepsilon_3$ depends on the cut-off $\Lambda$ and
on the masses of the pseudogoldstone, vector and axial vector bosons.
Except for small pseudogoldstone boson masses, it is negative.

Concerning hadronic colliders the discovery limit of charged
resonances through $WZ$ final state is lowered compared to $SU(2)$
case as soon as decay into pseudogoldstones bosons is
allowed~\cite{25}. If these particles can be copiously produced at LHC
they have to suffer a huge background since $P^0$ (resp $P^{\pm}$)
decay into $t\bar t$ or $b\bar b$ (resp $t\bar b$ and $\bar t b$).
These backgrounds have been studied for charged Higgs boson
discovery from $tb$ decays at LHC using SDC detector. It has been
shown that reasonably efficient and pure $b$ tagging is mandatory.
Our case deserves a careful study along the previous procedure to be
conclusive.

The most promising pseudogoldstone pair production mechanism at
$e^+e^-$ linear colliders is the resonant one through a $V^0$, which
is accompanied by an enhancement of $W$ pair production. The final
state to be considered is $t\bar b\bar t b$, already considered for
charged Higgs production.

\section{Conclusion}

We have studied the possibility of a strong interacting sector being
at the origin of the electroweak symmetry breaking, avoiding
elementary scalars. In absence of a specific definite theory of the
strong electroweak sector the BESS model provides for a rather general
frame based on custodial symmetry and gauge invariance.

A characteristic feature of the model is the occurence of spin one
resonances in the $TeV$ range and pseudogoldstone bosons in its
extended version. The idea of a strong electroweak sector is
quantitatively testable at LEP thanks to the recent precision
measurements: QCD scaled technicolor has already been excluded.

Linear $e^+e^-$ colliders are sensitive to the neutral vector
resonance $V^0$ whereas energetic hadronic colliders are well suited
to discover the charged ones. The characteristic feature of the
extended BESS model is the existence of pseudogoldstone bosons whose
detection needs a careful evaluation of backgrounds which remain to
be done.

\end{document}